\documentclass[final,5p,times,preprint]{elsarticle}

\usepackage{amsfonts}
\usepackage{amsmath}
\usepackage{amssymb}
\usepackage{textgreek}
\usepackage{bm}
\usepackage{comment}
\usepackage{xcolor}
\usepackage{booktabs}
\usepackage[normalem]{ulem}
\usepackage[export]{adjustbox}
\graphicspath{{./figures/}}
\journal{Physics Letters B}

\usepackage[utf8]{inputenc}
\usepackage[english]{babel}
\biboptions{comma,sort&compress}

\usepackage{amsmath}
\usepackage{graphicx}
\usepackage[colorinlistoftodos]{todonotes}
\usepackage[colorlinks=true, allcolors=blue]{hyperref}
\usepackage[normalem]{ulem} 

\newcommand{\pr}[1]{{\sc{\lowercase{#1}}}}

\def\etal{{\it et al. }}
\def\li#1#2{\ifnum#1=0 {$^{#2}$Li}
\else {$^{#1#2}$Li}
\fi
}
\def\lili#1#2#3#4{{{$^{#1#2,#3#4}$Li}}}
\def\bebe#1#2#3#4{{{$^{#1#2,#3#4}$Be}}}

\begin{document}

\begin{frontmatter}

\title{Evolution of two-neutrons configuration from $^{11}$Li to $^{13}$Li}
\author[cea]{P.~Andr\'e}
\author[cea]{A.~Corsi}
\cortext[mail]{Corresponding author}
\ead{acorsi@cea.fr}
\author[cea]{A.~Revel}
\author[rik,cns,tuda]{Y.~Kubota}
\author[FAMN]{J. Casal}
\author[FSU,FRIB,ANL]{K.Fossez}
\author[FAMN]{J. Gómez-Camacho}
\author[cea]{G. Authelet}
\author[rik]{H. Baba}
\author[tuda]{C. Caesar}
\author[cea]{D. Calvet}
\author[cea]{A. Delbart}
\author[cns]{M. Dozono}
\author[key]{J. Feng}
\author[ipno]{F. Flavigny}
\author[cea]{J.-M. Gheller}
\author[lpc]{J. Gibelin}
\author[cea]{A. Giganon}
\author[cea]{A. Gillibert}
\author[toh]{K. Hasegawa}
\author[rik]{T. Isobe}
\author[miy]{Y. Kanaya}
\author[miy]{S. Kawakami}
\author[ehw]{D. Kim}
\author[cns]{Y. Kiyokawa}
\author[cns]{M. Kobayashi}
\author[tod]{N. Kobayashi}
\author[toh]{T. Kobayashi}
\author[tit]{Y. Kondo}
\author[dae,rik,atom]{Z. Korkulu}
\author[tod]{S. Koyama}
\author[cea]{V. Lapoux}
\author[miy]{Y. Maeda}
\author[lpc]{F. M. Marqu\'es}
\author[rik]{T. Motobayashi}
\author[tod]{T. Miyazaki}
\author[tit]{T. Nakamura}
\author[kyo]{N. Nakatsuka}
\author[kyu]{Y. Nishio}
\author[cea,tuda]{A. Obertelli}
\author[kyu]{A. Ohkura}
\author[lpc]{N. A. Orr}
\author[cns]{S. Ota}
\author[rik]{H. Otsu}
\author[tit]{T. Ozaki}
\author[rik]{V. Panin}
\author[tuda]{S. Paschalis}
\author[cea]{E. C. Pollacco}
\author[tum]{S. Reichert}
\author[cea]{J.-Y. Rousse}
\author[tit]{A. T. Saito}
\author[kyu]{S. Sakaguchi}
\author[rik]{M. Sako}
\author[cea]{C. Santamaria}
\author[rik]{M. Sasano}
\author[rik]{H. Sato}
\author[tit]{M. Shikata}
\author[rik]{Y. Shimizu}
\author[kyu]{Y. Shindo}
\author[dae,rik,atom]{L. Stuhl}
\author[rik]{T. Sumikama}
\author[cea,tuda]{Y.L. Sun}
\author[kyu]{M. Tabata}
\author[tit]{Y. Togano}
\author[tit]{J. Tsubota}
\author[rik]{T. Uesaka}
\author[rik]{Z. H. Yang}
\author[kyu]{J. Yasuda}
\author[rik]{K. Yoneda}
\author[rik,key]{J. Zenihiro}

\address[cea]{D\'epartement de Physique Nucl\'eaire, IRFU, CEA, Universit\'e Paris-Saclay, F-91191 Gif-sur-Yvette, France}

\address[rik]{RIKEN Nishina Center, Hirosawa 2-1, Wako, Saitama 351-0198, Japan}

\address[cns]{Center for Nuclear Study, University of Tokyo, Hongo 7-3-1, Bunkyo, Tokyo 113-0033, Japan}

\address[FAMN]{Departamento de F\'{\i}sica At\'omica, Molecular y Nuclear, Facultad de F\'{\i}sica, Universidad de Sevilla, Apartado 1065, E-41080 Sevilla, Spain}
\address[FSU]{Florida State University, Tallahassee, Florida 32306, USA}
\address[FRIB]{FRIB Laboratory, Michigan State University, East Lansing, Michigan 48824, USA}
\address[ANL]{Physics Division, Argonne National Laboratory, Lemont, Illinois 60439, USA}
\address[tuda]{Department of Physics, Technische Universitat Darmstadt}
\address[pek]{Department of Physics, Peking University}
\address[ipno]{Institut de Physique Nucleaire Orsay, IN2P3-CNRS, F-91406 Orsay Cedex, France}
\address[lpc]{LPC Caen, ENSICAEN, Universite de Caen, CNRS/IN2P3, F-14050 Caen, France}
\address[toh]{Department of Physics, Tohoku University, Aramaki Aza-Aoba 6-3, Aoba, Sendai, Miyagi 980-8578, Japan}
\address[miy]{Department of Applied Physics, University of Miyazaki, Gakuen-Kibanadai-Nishi 1-1, Miyazaki 889-2192, Japan}
\address[ehw]{Department of Physics, Ehwa Womans University}
\address[tod]{Department of Physics, University of Tokyo, Hongo 7-3-1, Bunkyo, Tokyo 113-0033, Japan}
\address[tit]{Department of Physics, Tokyo Institute of Technology, 2-12-1 O-Okayama, Meguro, Tokyo 152-8551, Japan}
\address[atom]{MTA Atomki, P.O. Box 51, Debrecen H-4001, Hungary}
\address[kyo]{Department of Physics, Kyoto University, Kitashirakawa, Sakyo, Kyoto 606-8502, Japan}
\address[kyu]{Department of Physics, Kyushu University, Nishi, Fukuoka 819-0395, Japan}
\address[osa]{Research Center for Nuclear Physics, Osaka University, 10-1 Mihogaoka, Ibaraki, Osaka 567-0047, Japan}
\address[tum]{Department of Physics, Technische Universitat Munchen}
\address[dae]{Center for Exotic Nuclear Studies, Institute for Basic Science, Daejeon 34126, Republic of Korea}
\address[key]{School of Physics and State Key Laboratory of Nuclear Physics and Technology, Peking University, Beijing 100871, China}


\begin{abstract}
In this work we investigate the two-neutron decay of \li13 and of the excited states of \li11 populated via one-proton removal from $^{14}$Be and $^{12}$Be, respectively. 
A phenomenological model is used to describe the decay of \li11 and \li13. While the first one displays important sequential components, the second one appears dominated by the direct two-neutron decay. A microscopic three-body model is used to extract information on the spatial configuration of the emitted neutrons before the decay and shows that the average distance between the neutrons increases going from  \li11 to \li13.

\end{abstract}


\end{frontmatter}

\section{Introduction \label{sec:intro} }
The neutron-rich Lithium isotopes \lili1113 are known to be home to special features due to their extreme neutron over proton number imbalance. \li11, with very small one- and two-neutron separation energies of S$_n$=396~keV and S$_{2n}$=369~keV~\cite{nndc}, respectively, is a textbook case of a halo nucleus formed by a \li09 core plus two valence neutrons. Indeed, the phenomenon of neutron halo was revealed for the first time in \li11 via a total reaction cross section measurement \cite{tan85}. Besides being a halo nucleus, \li11 is a Borromean nucleus since \li10 is unbound. With a halo composed of two neutrons, a question naturally arises about the correlations developing among them. Several studies have been performed on \li11 dineutron correlations using different probes: transfer reaction \cite{tan08}, heavy-ions induced knockout reaction \cite{sim07}, Coulomb breakup \cite{nak06,mar00}, quasi-free scattering reactions \cite{kub20}, measurement of the dineutron decay from unbound excited states \cite{smi16}. All those results concur to some extent in the conclusion that the valence neutrons of \li11 sit in a spatially compact configuration called dineutron. \\
The unbound \li13 is much less studied, due to the fact that it is more difficult to reach experimentally. Its spectroscopy has been first reported in Ref.~\cite{aks08} and later in Ref.~\cite{koh13}. In both cases, \li13 is produced via one-proton removal from $^{14}$Be. While Ref.~\cite{aks08} observed a resonant state at 1.47(31)~MeV, the relative energy spectrum measured by Kohley \etal \cite{koh13} can be reproduced with a resonance at 120$^{+60}_{-80}$~keV. This discrepancy is most likely due to the limited two-neutron detection efficiency at small and large relative energy, respectively, and calls for further experimental studies.\\
In this manuscript, we present the study of the two-neutron decay of \lili1113 produced via one-proton removal from \bebe1214, respectively.  
Two-body (reduced) relative energy plots are interpreted using the Lednicky-Lyuboshitz model \cite{Led82} and a three-body model based on Ref.~\cite{JCasal19}. The second one is then used to extract information on the nature of the neutron-neutron correlations. A decay scheme is also proposed.

\section{The experiment \label{sec:exp} }
\subsection{Setup}
The experiment was performed using the Radioactive Isotope Beam Factory (RIBF) at the RIKEN Nishina Center. A primary beam of $^{48}$Ca at 345~MeV/nucleon with an intensity of 400~pnA underwent fragmentation on a Be target. The products of this reaction were then separated and identified in the BigRIPS fragment separator \cite{kubo03}. The resulting secondary beam was composed mainly of \li11 (70\%), $^{14}$Be (9\%) and $^{12}$Be (2.5\%), with a total intensity of 1$\times$10$^5$ pps. These nuclei arrived at the secondary target, a liquid hydrogen target with a thickness of 15~cm surrounded by the Time Projection Chamber (TPC) of the MINOS device \cite{obe14}, with an energy of 246, 265 and 340~MeV/nucleon, respectively. The reactions of interest here were $^{14}$Be($p$,2$p$)$^{13}$Li and $^{12}$Be($p$,2$p$)$^{11}$Li, followed by their respective decay $^{13}$Li$\rightarrow$ $^{11}$Li+2$n$ and $^{11}$Li$\rightarrow$ $^{9}$Li+2$n$. This work relies on the measurement of the energies and momenta of all the decay products, which was performed via the SAMURAI spectrometer \cite{kob13} using the vertex information provided by the MINOS TPC. The separation of the decay products was performed using the superconducting SAMURAI magnet, able to provide a magnetic field of 3.1~T, resulting in a bending angle of about 60°. The Li fragments were tracked thanks to two drift chambers, named FDC1 and FDC2, located between the secondary target and the SAMURAI magnet, and after the SAMURAI magnet, respectively. The charge and time-of-flight of these fragments were measured in two fragment hodoscopes (HODF), each one composed of 16 plastic scintillators with a total width of 160~cm. The neutrons were detected in the NEBULA array. This detector is composed of two walls, separated by 84~cm, made of two layers of 30 plastic scintillators each, making a thickness of 24~cm for each wall with a total area of 360$\times$180~cm$^{2}$.\\
Studying two-neutron (2$n$) decays implies carefully selecting the 2$n$ events out of all the events detected by the NEBULA array. An incident neutron can deposit energy in several plastic scintillators of the NEBULA array. As a result, a single neutron can be the origin of two or more signals in the NEBULA array, which can be wrongly analyzed as 2$n$ events. These events are called cross-talk events, and a procedure of cross-talk rejection was applied. This procedure is described in details by T.~Nakamura and Y.~Kondo in Ref.~\cite{naka16}, and it relies on conditions on the velocity and the time-of-flight of the signals in the NEBULA array. To test the performance of the chosen cross-talk rejection conditions, they have been applied on the data of a one-neutron (1$n$) decay channel, namely $^{10}$Li$\rightarrow$ $^{9}$Li+$n$, where $^{10}$Li is populated via a ($p$,$pn$) reaction on $^{11}$Li. For this 1$n$-decay channel, the survival rate of the 2$n$ events should be as low as possible. In this analysis, a survival rate of 2.3\% was obtained, which is comparable to the 2.9\% reported in Ref.~\cite{naka16}. Overall, when these cross-talk conditions are applied on the 2$n$ decay channel, the efficiency for detection of 2$n$ events, determined using a \pr{GEANT4} simulation, reaches a maximum of nearly 7\% at around 1~MeV (see Fig.~10 of Ref.~\cite{naka16}). The efficiency drops in the lower relative energy region due to the cross-talk rejection conditions, while the drop in the higher energy region is dominated by the geometrical acceptance of NEBULA. In total, 8275 $^{13}$Li decays and 7700 $^{10}$Li decays have been observed, after the rejection of cross-talk events.

\subsection{Experimental results \label{sec:exp}}
Three-body (3B) and two-body (2B) relative energies are calculated from the measured momenta using the following equations:
\begin{equation}
\begin{split}
&E_{fnn} = M_{fnn} - \left( m_{f} + 2m_{n} \right)~,\\
&E_{fn} = M_{fn} - \left( m_{f} + m_{n} \right)~,\\
&E_{nn} = M_{nn} - \left( 2m_{n} \right)~,
\end{split}
\end{equation}
with $m_{n}$ the neutron ($n$) mass and $m_{f}$ the fragment ($f$) mass, and with $M_{fnn}$ and $M_{fn}$, $M_{nn}$ the invariant mass of the 3B and 2B systems, respectively. 
These are defined as:
\begin{equation}
\begin{split}
&M_{i} = \sqrt{ E_{tot(i)}^{2}-\overrightarrow{P}_{tot(i)}^{2} }~,\\
\end{split}
\end{equation}
with $i= fnn, fn, nn$, $E_{tot}$ and $\overrightarrow{P}_{tot}$ the total energy and momenta of the system, respectively. One can also define the reduced 2B relative energies for the fragment+neutron and the neutron+neutron systems, noted $\varepsilon_{fn}$ and $\varepsilon_{nn}$, respectively, as:
\begin{equation}
\varepsilon_{fn} = \frac{E_{fn}}{E_{fnn}}, \varepsilon_{nn} = \frac{E_{nn}}{E_{fnn}}.
\end{equation}

Only events where neutrons are detected in different NEBULA walls are considered to build the respective experimental spectra. This event selection is motivated by the fact that the data obtained using neutrons detected in the same NEBULA wall are not sensitive to the low energy part of the relative-energy spectrum due to the cross-talk rejection procedure. The NEBULA response function for events with neutrons detected in different walls is applied consistently. The 3B spectra are tentatively reproduced using virtual states and/or resonances parameterized using equation (3) in Ref.~\cite{johansson2010unbound} adapted for a 3B system:
\begin{equation}
    \frac{d \sigma}{dE} \propto \frac{\Gamma_{l}(E_{fnn})}{(E_{res}+\Delta_{l}(E_{fnn})-E_{fnn})^{2}+\frac{1}{4} \Gamma_{l}(E_{fnn})^{2}} ,
\end{equation}
with $E_{res}$ the resonance energy, $l$ the orbital angular momentum, $\Gamma_{l}$ the width of the resonance and $\Delta_{l}$ the resonance shift. The definition of these quantities can be found in Ref.~\cite{johansson2010unbound}, as well as in Ref.~\cite{lane58}\footnote{equations (A.3a), (A.3b), (A12) and (A13)}. Although this is not a fully satisfactory way to extract the energy and width in case of a N-body resonance (N$>$2), this approach allows to conveniently fit experimental data with analytical functions. We believe its use is justified here since we do not extract the resonance widths, and the shift on the centroid energies is negligible at energies of a few hundreds keV, where narrow peaks appear.
The 2B spectra are reproduced with functions issued from the phenomenological three-body decay model described hereafter. Those theoretical functions are folded with the response function of the setup via a \pr{GEANT4} simulation.

\subsubsection{Phenomenological three-body decay model}
The two neutrons emitted during the three-body decay can be considered free and independent of each other, or correlated. In order to include the different correlations observed above a pure phase-space distribution of events, we use the model developed in Ref.~\cite{Mar01}. This model does not include the microscopic structure of the initial state and treats the effects of Final State Interaction (FSI) and resonances on the fragment+2$n$ phase-space decay phenomenologically (see Ref.~\cite{Lau19} for more details on its applicability). In brief, the experimental 3B relative energy distribution is used to generate events with $\Vec{p_{f}}$, $\Vec{p_{n_1}}$, $\Vec{p_{n_2}}$ following either the three-body phase space (direct decay), or twice the two-body phase space through a fragment-$n$ resonance (sequential). In the latter case, a neutron and the fragment-$n$ resonance are generated first, followed by the decay of the resonance. In both cases, the $n$-$n$ FSI is introduced via a probability P(q$_{nn}$) with the form of the $n$-$n$ correlation function \cite{Led82}, which depends on the space-time parameters (r$_{nn}$,$\tau$) of a Gaussian two-neutron source. For a given resonance observed in the three-body relative energy spectrum, a fit of the experimental two-body reduced relative energy spectra can be performed using a combination of a direct decay component and $N$ sequential decay components corresponding to $N$ fragment-$n$ resonances, leaving us with 4$N$+1 free parameters: the neutron source space parameter r$_{nn}$ , the energy E$_{r_i}$ and width $\Gamma_{r_i}$ of the fragment-$n$ resonance, the lifetime of the intermediate fragment-$n$ system $\tau_{i}$ and the fraction of sequential decay $\alpha_{r_i}$ with i=1,...,$N$. The number of free parameters is further reduced by linking the lifetime of the fragment-$n$ resonance and the delay induced in the neutron emission using $\tau_{i}=\hbar c/\Gamma_{r_i}$, as done in Ref.~\cite{Lau19}.

\subsubsection{$^{13}$Li: results and discussion \label{sec:li13}}
\begin{figure}[h!]
    \centering
    \hspace*{-1cm}
\includegraphics[height=15cm]{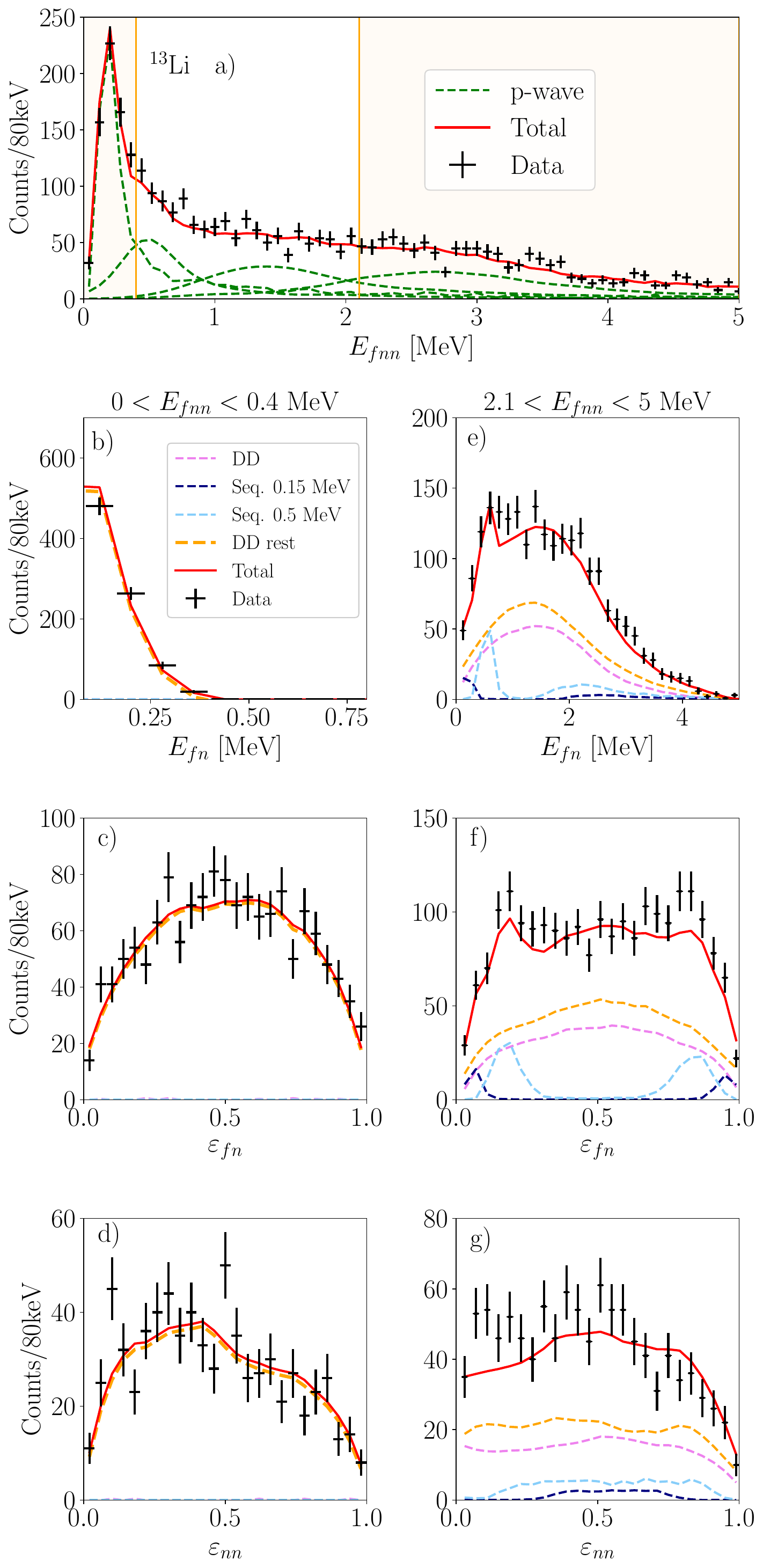} 
    \caption{Fits of the 3B, 2B and reduced energy spectra of $^{13}$Li. "DD" relates to the direct decay of the 2.8 MeV resonance, and "DD rest" relates to the direct decay of the other resonances. "Seq. 0.15 MeV" and "Seq. 0.5 MeV" relate to the sequential decay, via the 0.15 MeV and 0.5 MeV resonances, respectively, in $^{12}$Li. a) \li13 3B relative energy spectrum;  b) \li11+$n$ 2B relative energy spectrum  gated on $ 0 <E_{fnn}< 0.4$~MeV; c) 2B \li11-$n$ reduced relative energy spectrum, same gate as b); d) 2B $n$-$n$ reduced relative energy spectrum, same gate as b); e) \li11+$n$ 2B relative energy spectrum  gated on $ 2.1 <E_{fnn}< 5$~MeV; f) 2B \li11+$n$ reduced relative energy spectrum, same gate as e); g) 2B $n$-$n$ reduced relative energy spectrum, same gate as e). The slices of 3B relative energy spectra presented here are shaded in panel a).}
    \label{fig:13Li_3B_2B}
\end{figure}

The 3B energy spectrum of \li13 (\li11+$n$+$n$) is shown in Fig.~\ref{fig:13Li_3B_2B}~a). It can be reproduced in a satisfactory way with the overlap of 4 resonant structures centered at 0.16(1), 0.45(6), 1.47~\cite{aks08} and 2.8(2)~MeV. Note that only the first resonance can be identified unambiguously and its energy is consistent with the one of the proposed ground state reported in Ref.~\cite{koh13}. The third resonance energy has been fixed using the results from Ref.~\cite{aks08}, while the remaining ones are fitted in order to reproduce the spectrum using the minimum number of resonances. We would like to stress that our data set does not allow to produce an unambiguous decomposition of the 3B energy spectrum, but only a convenient choice for the following analysis of the \li13 decay.\\
Given the absence of visible structures in the \li11+$n$ 2B relative energy spectrum, we initially assume that \li13 decays directly to \li11 via two-neutron emission. We compare the 2B and reduced 2B relative energy plots for 4 slices of the 3B relative energy (E$_{fnn}$) spectrum centered around each dominating resonance ($0 <E_{fnn}< 0.4$~MeV, $0.4 <E_{rel,3B}< 1$~MeV, $1 <E_{fnn}< 2.1$~MeV, $2.1 <E_{fnn}< 5$~MeV) with a calculation using two-neutron correlation functions modeled via the Lednicky-Lyuboshitz formalism. Two of those slices are shown in Fig.\ref{fig:13Li_3B_2B}~b)-d) and e)-g), respectively. The 2B (reduced) energy spectra are satisfactorily reproduced for all slices using only one direct component, except for the 2.1~MeV$<E_{fnn}<$5~MeV slice. Here some typical features of sequential decay appears, as can be better seen in Fig.~\ref{fig:13Li_3B_2B} f). Scanning the \li11+$n$ spectrum, we find a slice corresponding to 2.76 $< E_{fnn} <$ 3.36~MeV where distinct features appear, which are interpreted as excited states of \li12 at 0.15$^{+0.05}_{-0.01}$~MeV and 0.50$^{+0.1}_{-0.01}$~MeV, consistent with the results of Ref.~\cite{koh13}. Our dataset does not allow us to investigate the presence of the s-wave virtual state observed in $^{12}$Li by the authors of Ref. \cite{aks08}, since its contribution in the $\epsilon_{fn}$ spectrum will be indistinguishable from the direct decay. The addition of a sequential component in the decay scheme (already implemented in Fig. \ref{fig:13Li_3B_2B}~e)-g)) improves the agreement with the 2B (reduced) relative energy plots in 2.1 $< E_{fnn} <$ 5~MeV slice. 

\subsubsection{$^{11}$Li: results and discussion \label{sec:li11}}
\begin{figure}[h!]
    \centering
\includegraphics[height=13cm]{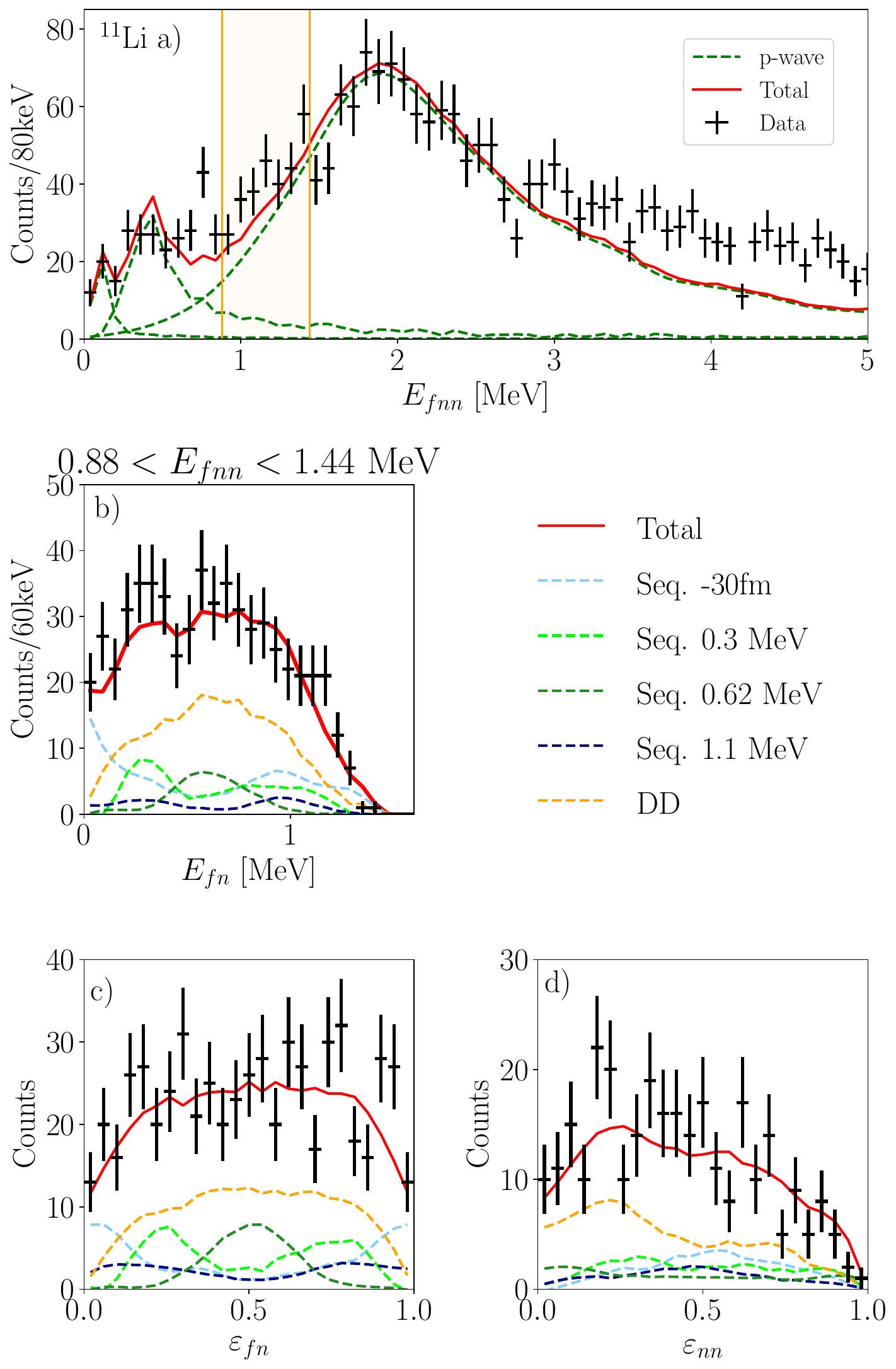} 
    \caption{Fits of the 3B, 2B and reduced energy spectra of $^{11}$Li. "Seq. -30fm", "Seq. 0.3 MeV", "Seq. 0.62 MeV" and "Seq. 1.1 MeV" relate to the sequential decay, via the virtual state with $a_{s}=$-30 fm, the 0.3 MeV resonance, the 0.62 MeV resonance and the 1.1 MeV resonance, respectively, in $^{10}$Li. "DD" relates to the direct decay of the 2.13 MeV resonance in $^{11}$Li. a)\li11 3B relative energy spectrum;  b) \li9+$n$ 2B relative energy spectrum  gated on $ 0.88 <E_{fnn}< 1.44$~MeV;  c) 2B \li9+$n$ reduced relative energy spectrum for the same gate as b); d) 2B $n$+$n$ reduced relative energy spectrum for the same gate as b). The slice of 3B relative energy spectrum presented here is shaded in panel a).}
    \label{fig:11Li_3B_2B}
\end{figure}

We repeat the same kind of analysis for \li11. The full 3B (\li9~+$n$+$n$) spectrum cannot be completely reproduced using a limited amount of structures, especially for relative energies above 3.5 MeV, possibly due to the approximation in the resonances parametrization. We adopted 3 resonant structures centered at 0.08(2)~MeV, 0.39(6)~MeV, and 2.1(3)~MeV, where peaks appear in the spectrum. The second resonance is consistent with a previous measurement from Ref.~\cite{nak06}. Looking at Fig.~\ref{fig:11Li_3B_2B}~a), we can see that the data are not well reproduced on the tails of the 2.1~MeV resonance, which may hide several structures not distinguishable here, as in the previous work shown in Ref. \cite{zin97} who adopted a similar resonant state. Differently from \li13, in this case a sequential component clearly appears in the \li9~+$n$ 2B relative energy spectrum. We compute the 2B relative energy spectrum for three representative regions of the 3B one and adjust with a calculation using two-neutron correlation functions modeled via the Lednicky-Lyuboshitz formalism. One is centered around the low energy resonances ($ 0 <E_{fnn}< 0.48$~MeV), and another around the high energy resonance ($ 1.92 <E_{fnn}< 2.72$~MeV). The intermediate region ($ 0.88 <E_{fnn}< 1.44$~MeV) is centered around an area displaying an excess of counts, which could be explained by an additional resonance. A constraint is to use the minimum amount of resonances, and it appears that, in this range, the other spectra can be reproduced with the initial three resonances, as shown in Fig. \ref{fig:11Li_3B_2B}. 
Fig.~\ref{fig:11Li_3B_2B} b) shows the 2B relative energy spectrum of \li10 for the $ 0.88 <E_{fnn}< 1.44$~MeV interval. This spectrum is fitted with the overlap of four resonant structures corresponding to four intermediate states in \li10: a virtual state with scattering length fixed at a$_s$=-30~fm~\cite{sim07}, and three fitted resonances centered at 0.30(6)~MeV, 0.62(10)~MeV and 1.1(1)~MeV. The existence of those states is postulated based on the features of the 2B relative energy spectrum observed gating on the $ 0.88 <E_{fnn}< 1.44$~MeV (shown in Fig. \ref{fig:11Li_3B_2B}~b)-d)) and $1.92 <E_{fnn}< 2.72$~MeV intervals. The 0.3~MeV and 0.62(10)~MeV resonance are compatible with a measurement shown in Ref.~\cite{zin97}. The 1.1~MeV resonance is compatible with a measurement shown in Ref.~\cite{smith15}. The same functions are used to reproduce the reduced 2B relative energy spectra in Fig.~\ref{fig:11Li_3B_2B}.

\subsubsection{Summary of experimental results}
\begin{figure}[h!]
    \centering
\includegraphics[height=5cm]{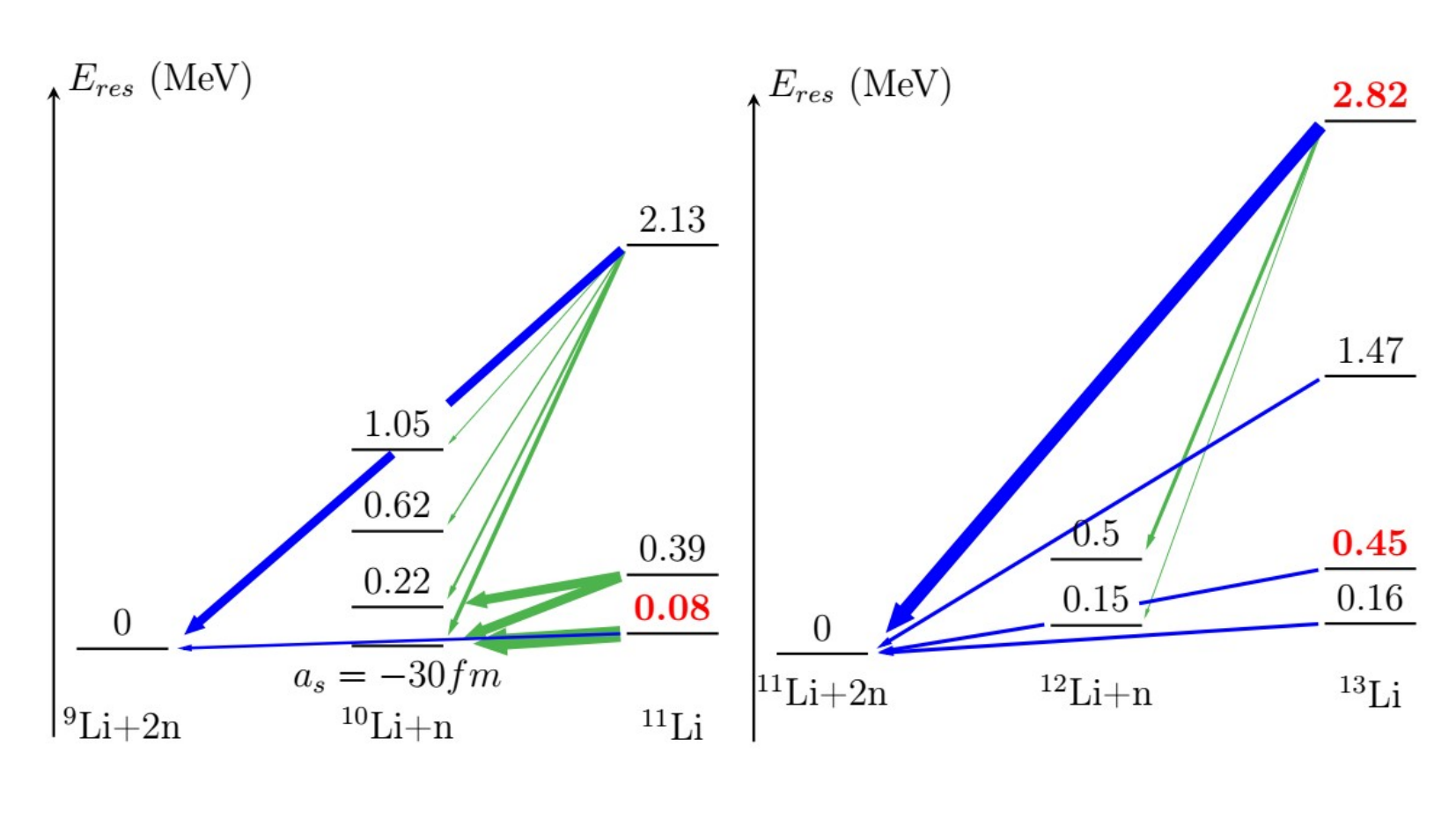} 
    \caption{Level schemes for \li11 (left) and \li13 (right). The arrows indicate the type of decay. The green ones indicate sequential decays, and the blue ones indicate direct decays. The levels in bold and red font are the ones that have been postulated in this work.}
    \label{fig:LevelSchemes}
\end{figure}

Based on the results presented in Sect. \ref{sec:exp}, we built a decay scheme for the two Li isotopes of interest that is presented in Fig. \ref{fig:LevelSchemes}. The probability of each decay path is represented by the width of the arrow representing the transition. The uncertainty on this probability is typically 10-15\% in the case of \li11, and 5\% in the case of \li13. In the case of \li13, we observe sequential decay through \li12 only from the 3B relative energy region between 2 and 5~MeV. Otherwise, the decay is mainly direct in agreement with the findings of Ref.~\cite{koh13}. Conversely, the decay of \li11 is dominated by sequential decay via \li10 intermediate states.

\section{Comparison with theoretical calculations \label{sec:theo} }

The relative-energy distributions of the two-neutron decay can be described naturally within three-body models \cite{SWang17, Grigo18}. 
This requires, first, a realistic description of the $\text{core}+n+n$ decaying state, and second, a formalism that provides the corresponding angular and energy correlations in the final state. In this work, we adopt the method proposed in Ref.~\cite{JCasal19} and already applied in Ref.~\cite{mon24}. The main aspect of the method is to define a resonance operator, the eigenstates of which describe localized continuum structures as a combination of discretized continuum states of different energy. While lacking the proper asymptotic behaviour, it was shown that a discrete-basis representation is enough to describe the resonance energy and decay width reasonably, as well as the initial-state dineutron correlations.

Here we focus on the decay properties of the $^{13}$Li ground-state resonance. The corresponding $^{11}\text{Li}+n+n$ wave function was built within the hyperspherical description~\cite{JCasal18,JCasal19,Zhukov93}. For simplicity, the spin of the core was ignored, and we used as a starting point the binary potentials employed in Ref.~\cite{gomezramosplb17} for $^{11}$Li calculations. The $\text{core}+n$ potential was adjusted so that $^{12}$Li presents a p$_{1/2}$ resonance around 0.4 MeV above the $^{11}$Li+n threshold, close to the experimental levels in Fig. \ref{fig:LevelSchemes}. With these ingredients, the ground-state resonance of $^{13}$Li at 0.2 MeV has a sizeable width of 0.1 MeV, which is somewhat compatible with the experimental width of 0.16 MeV observed for the lowest energy structure. No excited states is predicted within this three-body model. 
Then, we find the outgoing solution of an inhomogeneous equation, where the source term is obtained from the square-normalizable state describing the resonance and takes into account the interactions~\cite{casaltbd}. The amplitudes in terms of asymptotic functions provide the corresponding relative-energy (or momentum) distributions.

\begin{figure}[h!]
    \centering
\includegraphics[height=4cm]{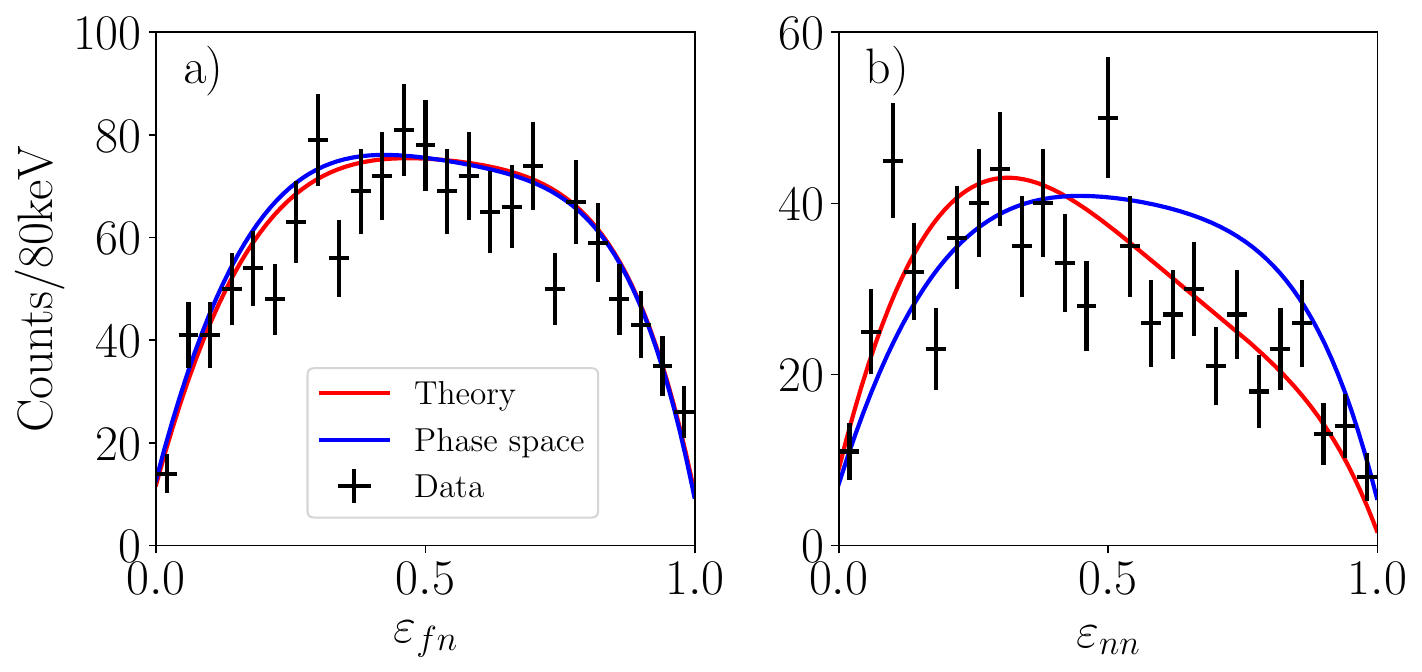} 
    \caption{Comparison of the experimental $\varepsilon_{fn}$ and $\varepsilon_{nn}$ distributions with the theoretical three-body calculations for the decay of \li13 ground-state resonance (red line). Experimental data are the same as in Fig. \ref{fig:13Li_3B_2B} (c,d). The phase space distribution, without \textit{n-n} correlations, is also shown as a reference (blue line).}
    \label{fig:13Li_th}
\end{figure}

The results for the reduced relative-energy spectra are presented in Fig.~\ref{fig:13Li_th} together with the experimental data (from Fig.~\ref{fig:13Li_3B_2B} panels c) and d)). 
Theoretical calculations have been used as input of the simulation described in Sec.~2.1 to take into account the experimental resolution and acceptance. The theoretical lines capture the general trend of the data for both $\varepsilon_{fn}$ and $\varepsilon_{nn}$ distributions, which is consistent with a direct decay for the ground-state resonance. However, the model does not predict any other resonances that exhibit signatures of sequential decay, as observed in the data for the $ 2.1 < E_{rel,3B} < 5 $ MeV slice. A better understanding of the decay scheme may require a more refined description of $^{13}$Li, in particular regarding the $\text{core}+n$ interaction and the possible effect of core-excited states. 

\begin{figure}[h!]
    \centering
 \includegraphics[width=1\linewidth]{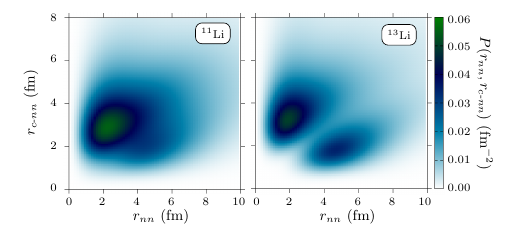} 
 
     \caption{Probability density distribution for the ground state of \li11 (left) and \li13 (right) from the three-body calculation.}
    \label{fig:density}
\end{figure}

Figure \ref{fig:density} shows the probability density obtained within the three-body model as a function of the neutron-neutron and core-neutrons distance. For comparison, the result for the ground state of the two-neutron halo nucleus $^{11}$Li, adopting the structure model described in Refs.~\cite{gomezramosplb17,casalplb17}, is also shown. 
The case of \li11 presents a rather diffuse spatial distribution around the dineutron-like structure, while for \li13 the distribution splits in two parts. 
The RMS $r_{n\text{-}n}$ distance in the present three-body calculations is 6.5 fm for the ground state of $^{11}$Li, 
and 8.1 fm for that of $^{13}$Li. Overall, the increase from $^{11}$Li to $^{13}$Li suggests a more dilute spatial configuration of the neutrons of \li13, as expected for a low-lying resonant ground state. It is worth noting that a five-body description of $^{13}$Li, i.e., four neutrons around a $^9$Li core, would provide a more consistent description of $^{11,13}$Li, but such a model for the resonance decay is not yet available.

\section{Conclusions \label{sec:end} }
We have presented here a study of the two-neutron decay of \li13 and \li11 excited states. Experimental data on two-neutron decay have been first interpreted within a phenomenological approach based on the formalism developed in Ref. \cite{Led82} that allowed to pinpoint a sequential contribution not yet observed in \li13 decay to \li11 + $2n$. Overall, the decay of \li13 is dominated by the so-called direct two-neutron decay contribution, while the one of \li11 presents important sequential components. Experimental spectra of \li13 have been also compared to the ones obtained with a three-body model calculations for the decay \cite{JCasal19,casaltbd}. The shape of the fragment-neutron and neutron-neutron reduced energy distributions, capturing the direct nature of the decay, is well reproduced. The three-body model predicts a more dilute spatial distribution of the neutrons when going from \li11 to \li13. 
Overall, we have shown that the two-neutron decay displays a sensitivity to neutron-neutron correlations. Even if the present data constitute a step forward in terms of statistics and resolution, complementary measurements and analyses may bring additional information on the partial-wave composition of the relative energy spectrum. One may consider, for example, performing a multiple decomposition analysis of the missing momentum spectrum. Our results encourage an improvement of the theoretical description of the excitation spectrum and an investigation of the dependence of the observed properties (energy spectrum, partial wave content) on the reaction mechanism employed. 

\section*{Acknowledgements}
This work has been supported by the European Research Council through the ERC Starting Grant No. MINOS-258567. 
J.C.\ acknowledges financial support by MCIN/AEI/10.13039/ 501100011033 under I+D+i project No.\ PID2020-114687GB-I00 and by the European Union's Horizon 2020 research and innovation programme under the Marie Skłodowska-Curie grant agreement No.\ 101023609.
J.G., F.M.M. and N.A.O.\ acknowledge partial support from the Franco-Japanese LIA-International Associated Laboratory for Nuclear Structure Problems as well as the French ANR14-CE33-0022-02 EXPAND. Z.K. and L.S. acknowledge partial support by the Institute for Basic Science (IBS-R031-D1). S.P. acknowledges the support of the UK STFC under contract numbers ST/L005727/1 and ST/P003885/1 and the Deutsche Forschungsgemeinschaft (DFG, German Research Foundation) Project-ID 279384907 - SFB 1245.
KF acknowledges the suport of the U.S. Department of Energy, Office of Science, Office of Nuclear Physics, under the FRIB Theory Alliance award no.\ DE-SC0013617.

\bibliography{biblio}

\end{document}